# ZIF-90 treats fungal keratitis by promoting macrophage apoptosis and inhibiting inflammatory response

Xueyun Fu[1], Jing Lin[1], Qian Wang[1], Lina Zhang[1], Ziyi Wang[1], Menghui Chi[1], Daohao Li[2*], Guiqiu Zhao[1*], Cui Li[1*]

[1] Department of Ophthalmology, The Affiliated Hospital of Qingdao University, Qingdao, People's Republic of China

[2] State Key Laboratory of Bio-fibers and Eco-textiles, Institute of Marine Biobased Materials, College of Materials Science and Engineering, Qingdao University, Qingdao

Correspondence authors:

E-mail addresses: lidaohao@qdu.edu.cn (D. Li), zhaoguiqiu_good@126.com (G. Zhao), yankelicui@126.com (C. Li)

# Abstract

**Background:** Fungal keratitis is a severe vision-threatening corneal infection with a prognosis influenced by fungal virulence and the host's immune defense mechanisms. The immune system, through its regulation of the inflammatory response, ensures cells and tissues can effectively activate defense mechanisms in response to infection and injury. However, there is still a lack of effective drugs that attenuate fungal virulence while relieving the inflammatory response caused by fungal keratitis. Therefore, finding effective treatments to solve these problems is particularly important.

**Methods:** We synthesized ZIF-90 by water-based synthesis and characterized by SEM, XRD, XPS, and FT-IR. In vitro experiments included CCK-8, spore adhesion test, MIC, biofilm, mycelium electron microscopy, PI staining, calcium fluoride white staining, apoptosis staining, PCR, and ELISA. These evaluations verified the disruptive effects of ZIF-90 on *Aspergillus. fumigatus* spore adhesion, morphology, cell membrane, cell wall, biofilm, and the influence of pro-inflammatory factors TNF-α and IL-6, and the effect of ZIF-90 on apoptosis. In vivo, corneal toxicity test, establishment and treatment of mycotic keratitis mouse model, HE staining of the cornea, and immunofluorescence staining were used to evaluate the efficacy of ZIF-90 in the procedure of fungal keratitis. In addition, to investigate whether the metal-ligand zinc and the organic ligand imidazole act as essential factors in ZIF-90, we investigated the in vitro antimicrobial and anti-inflammatory effects of ZIF-8, ZIF-67 and MOF-74 (Zn) by MIC and ELISA experiments.

**Results:** ZIF-90 has therapeutic effects on fungal keratitis, which could break the protective organelles of *Aspergillus. fumigatus*, such as the cell wall. In addition, ZIF-90 can avoid excessive inflammatory response by promoting apoptosis of inflammatory cells. The results demonstrated that both zinc ions and imidazole possessed antimicrobial and anti-inflammatory effects. In addition, ZIF-90 exhibited better biocompatibility compared to ZIF-8, ZIF-67, and MOF-74 (Zn).

**Conclusion:** ZIF-90 has anti-inflammatory and antifungal effects and preferable biocompatibility, and has great potential for the treatment of fungal keratitis.

**Keywords**： zeolite imidazolate framework, nanomedicine, fungal infection, keratitis

# 1. INTRODUCTION

Fungal keratitis is a serious blinding eye disease that can lead to permanent vision loss.[1,2] Aspergillus is one of the main causative agents of fungal keratitis.[3] Fungal keratitis can occur due to a variety of reasons, such as trauma, wearing corneal contact lenses, and over usage of glucocorticoids. It is estimated that more than one million patients worldwide suffer from fungal keratitis each year, and about three-quarters of them are blind due to the disease.[2] Fungal keratitis has become a major public health problem in many countries, especially in developing countries. How to effectively and economically treat fungal keratitis remains a challenge.

Reducing the virulence of the fungus and controlling the host's defenses are key to the treatment of fungal keratitis, and this critical issue determines the prognosis of fungal keratitis. As to the fungal invasion, the host immune cells can limit the tissue damage caused by fungal infection. Bauer et al. proved that subconjunctival macrophages are the first line of defense cells in the cornea, which promotes the secretion of inflammatory cytokines and recruits a variety of immune cells to help the host fight fungal infection.[4-6] Nevertheless, the overactivation of macrophages in the cornea leads to an excessive immune response. The excessive immune response brings about an uncontrollable inflammatory response that destroys the corneal stroma.[7,8] Thus, controlling the inflammatory response is particularly important in the treatment of fungal keratitis. Unfortunately, current therapeutic agents are unable to control the excessive inflammatory response, which poses a potential problem in the treatment of fungal keratitis. Moreover, these drugs have poor permeability and definite corneal toxicity.

ZIF-90, as a kind of MOF material, has the advantages of excellent biocompatibility and low cytotoxicity.[9,10] In recent years, nano-delivery systems have been widely used in treating eye diseases due to their low toxicity and efficient delivery. ZIF-90, as a member of the nanocarrier, has a wide range of applications in targeted drug delivery, such as the delivery of metronidazole as a drug delivery agent for the treatment of sepsis.[11] These broad applications highlight the versatility and potential of ZIF-90 as a versatile material.[12-14] However, the immunogenicity and efficacy of ZIF-90 in the treatment of fungal keratitis remain unclear. ZIF-90 is composed of metal ions zinc and imidazole ligands. Zinc ion has been proven to have anti-inflammatory and antibacterial effects.[15-17] Azole drugs are a class of classical antifungal drugs widely used in clinical settings,[18,19] so this paper predicts that ZIF-90 may have therapeutic effects on fungal keratitis.

In this study, ZIF-90 was synthesized and characterized by the water-based synthesis method. Then, the anti-inflammatory and antifungal effects of ZIF-90 were studied using the spore adhesion test, ELISA, and other tests. We found that in vitro, ZIF-90 can inhibit the growth of *Aspergillus. fumigatus* (*A. fumigatus)* by destroying the cell wall cell membrane, reducing spore adhesion to HCECs, and controlling inflammatory responses by promoting macrophage apoptosis. In vivo, ZIF-90 can treat fungal keratitis by reducing the infiltration of inflammatory cells in the mouse cornea and reducing the number of inflammatory factors. Finally, the antibacterial and anti-inflammatory effects of ZIF-8, ZIF-67, and MOF-74 (Zn) were verified in vitro, proving that both zinc ions and imidazole groups contained in ZIF-90 can inhibit the growth of *A. fumigatus* and have anti-inflammatory effects. The presence of aldehyde groups on the imidazole group can improve the biocompatibility of ZIF-90.

## 2. Materials and methods

**Synthesis and characterization of ZIF-90**

A zinc nitrate aqueous solution was prepared as solution A, while imidazole-2-carboxaldehyde (ICA) and polyvinylpyrrolidone (PVP) were dissolved to form solution B. In the water/ethanol/PVP system, zinc nitrate was dissolved in an ethanol medium to form solution A. The amount of PVP and $Zn^{2+}$ to ICA ratio was meticulously controlled. Solutions A and B were mixed at room temperature for 30 min. ZIF-90 was collected by centrifugation and thoroughly washed with methanol. The powder was vacuum-dried at 50 °C.[20]

The shape and appearance of the samples were examined using a scanning electron microscope (SEM; JSM-7001F; JEOL, Tokyo, Japan), and the particle size of the samples was analyzed using Image J software. X-ray diffraction (XRD) analysis was conducted by using a Rigaku diffractometer (Japan), and employing x-ray radiation across a 5°-40° range, with a scanning rate set at 5° $min^{-1}$. X-ray photoelectron spectroscopy (XPS) was carried out using a VG Scientific ESCALab220i-XL spectrometer equipped with a monochromatic Al Kα X-ray source. Fourier Transform Infrared Spectrometer (FT-IR; Themo, USA) was used to measure the samples in the wavelength range of 1500 to 2500 $cm^{-1}$.

### Acquisition of *A. fumigatus* hyphae and spores

*A. fumigatus* spores (Strain 3.0772, China General Microbiological Culture Collection Center) were placed in an aseptic Salter medium and cultured in a shaking table for 5-7 days (300 rpm, 37 °C). When the spores grew into hyphae and formed into balls, they were ground with a sterile grinding rod. After the hyphae were ground evenly, they were washed with Phosphate buffer saline (PBS; Solarbio). Then, the hyphae were soaked with 75% ethanol and placed in a refrigerator at 4 °C overnight for inactivation. Upon inactivation, the hyphae were washed with PBS and diluted with a high-sugar culture solution to a concentration of $1×10^8$. The spores were cultured on sabouraud agar medium (SDA) (Hopebio, China) and washed with PBS. The spore suspension was diluted to a suitable concentration during usage.

### Human corneal epithelial cells (HCECs) fungal spore adhesion assay

HCECs were inoculated and cultured in a cell incubator following standard protocols.[21]

ZIF-90 (64 μg/mL) and spore suspension (1×10$^7$ CFU/mL) were added when the cells grew into 30%-50% of the plate pore area. HCECs and spores were cultured for 3 h, and the culture medium was sucked out. Then, HCECs and spores were washed 3 times with sterile PBS and then fixed with methanol, next the methanol was sucked out and dried, HCECs were stained by HE and observed spore adhesion under a microscope (Axiovert; Zeiss, Jena, Germany, 400×).

**Cell viability test**

The Cell Counting Kit 8 (CCK-8) kit was used to detect the cytotoxicity of ZIF-8, ZIF-67, MOF-74 (Zn), and ZIF-90 against HCECs and RAW 264.7 cells. RAW 264.7 cells (Obtained from the Academy of Sciences in Shanghai) were inoculated onto 96-well plates. When the cells grew into 80% of the pore plate area, different concentrations (2, 4, 8, 16, 32, 64, 128 μg/mL) of ZIF-8, ZIF-67, MOF-74 (Zn) and ZIF-90 were added. The four MOF materials were added to cck-8 (HY-K0301) post 24 h of co-culture with cells, and the absorbance at 450 nm was measured after 1 h of culture.

**Minimum inhibitory concentration (MIC)**

A spore suspension of 3×10$^5$ CFU/mL was inoculated into a 96-well plate, followed by the addition of ZIF-8, ZIF-6, and ZIF-90 to achieve final concentrations of 2, 4, 8, 16, 32, 64, and 128 μg/mL, respectively. The 96-well plate was then incubated in a constant temperature incubator (37 °C, 5% CO$_2$) for 24 h to allow for co-cultivation of the compounds with the spores. The absorbance at 540 nm was measured as soon as the incubation period.

**Biofilm Inhibition Experiment**

The spore suspension of 3×10$^5$ CFU/mL and ZIF-90 of different concentrations (4, 8, 16, 32, 64, 128 μg/mL) were added to the 24-well plate. The spore and ZIF-90 were cultured in a constant temperature incubator (37 °C, 5% CO$_2$) for 24 h until thin hyphae grew. Then, the surface hyphae were removed with an insulin needle, and the sand culture medium was extracted. The biofilm in the bottom of the 24-well plate was washed with PBS 3 times and dried at room temperature. Once drying, the 99% methanol was added to the 24-well plate to fix the biofilm for 20 min. Succeeding removing the methanol, the biofilm was rinsed with sterile deionized water and dried at room temperature. Crystal violet stain was added to the plate and soaked at room temperature. After 15 min, the dye solution was removed, and the unbound crystal violet was fully washed with sterilized PBS. Then, the biofilm was fully decolorized with 95% ethanol. The mixture of ethanol and crystal violet was absorbed and transferred to a new 96-well plate, and the optical density was measured at 570 nm.

**Propidium Iodide (PI)**

The spore suspension of $3\times10^5$ CFU/mL was inoculated with a 6-well plate, and a layer of hyphae was formed later on incubation in a constant temperature incubator (37 °C, 5% $CO_2$) for 24 h. Different concentrations (32 and 64 μg/mL) of ZIF-90 were covered on the surface of hyphae. The incubation was continued for 24 h, and then the culture solution was sucked out. The hyphae were washed 3 times with sterile PBS and then stained with PI staining solution (Solarbio, Beijing, China). Then, the hyphae were incubated for 15 min and protected from light. The hyphae were photographed with a fluorescence microscope (Nikon, Tokyo, 200×).

**Calcofluor White Stain (CFW; Sigma)**

A spore suspension of $3\times10^5$ CFU/mL was co-cultured with ZIF-90 in a 12-well plate for 24 h. The 12-well plate was washed 3 times with sterile PBS. The Calcium Fluoride White Stain was added to the 12-well plate. Then, the plate was incubated for 15 min at room temperature, avoiding light. The fluorescence intensity was observed under a fluorescence microscope (Nikon, Tokyo, 200×) and photographed.

**Hyphae scanning electron microscopy (SEM)**

A $3\times10^5$ CFU/mL spore suspension was inoculated with a 6-well plate. A layer of hyphae was formed after 24 h of incubation in a constant temperature incubator (37 °C, 5% $CO_2$), and ZIF-90 (64 μg/mL) was covered on the surface of the hyphae. The incubation was continued for 24 h. Next, the hyphae were washed 3 times with sterile PBS and fixed with 2.5% glutaraldehyde solution. The hyphae morphology was observed under a scanning electron microscope (JSM-840; JOEL, 200×).

**Cell culture and hyphae stimulation**

RAW 264.7 cells were cultured in DMEM high sugar medium (Biological Industries, Israel) supplemented with 10% fetal bovine serum. RAW 264.7 cells in cell culture plates were stimulated with *A. fumigatus* filaments for 1 h. Then ZIF-90 was added, and the stimulation continued for 8 h or 24 h. Cells were incubated in a constant temperature incubator (37 °C, 5% $CO_2$).

**Annexin V-FITC/ PI Apoptosis staining (E-CK-A211, Elabscience)**

RAW 264.7 cells were inoculated in 12-well plates, and subsequently the cell density grew to $1\times10^5$, inactivated hyphae were added to stimulate the cells for 1 h. Afterward, ZIF-90 (64 μg/mL) was added to treat the cells for 24 h. Cells were collected and suspended with 500 of diluted 1×Annexin V Binding Buffer. 5 μL of Annexin V-FITC Reagent and 5 μL of PI Reagent were added to the cell suspension, and the cells were incubated at room temperature and protected from light for 15-20 min then were detected by a flow cytometer (Agilent NovoCyte 2060R) within 1 h.

**Apoptosis Fluorescence Hoechst Staining (C0021, Solarbro)**

RAW 264.7 cells were inoculated in 12-well plates. When the cell density reached $1\times10^5$, inactivated hyphae were added to stimulate for 1 h. Then, ZIF-90 (64 μg/mL) was added to continue the incubation for 8 h. The 12-well plate was washed with PBS, and then a PBS and Hoechst staining solution were added. The 12-well plate was placed in the refrigerator at 4 °C for 20-30 min. The photographs of stained cells were taken under a fluorescence microscope (EVOS M5000, Thermo; 400×).

**Corneal toxicity test**

The toxicity of ZIF-90 to the mouse cornea was investigated based on an ocular toxicology study (Driaze Toxicity Assay).[22] Mice were treated with ZIF-90 (5 μL once a day; 0.1% DMSO, 64 and 128 μg/mL) in the right eye, and the left eye served as a blank control. Fluorescence staining (CFS) of the corneas of mice on days 1, 3, and 5 were observed under cobalt blue light to reflect the toxicity of ZIF-90 to the mouse cornea. The area and density of corneal ulcers, iritis, and conjunctival erythema were scored.

**Establishment and treatment of mice keratitis model**

Mice (female C57BL/6 mice 6-8 weeks, 20-30 g weight) were narcotized with 8% chloral hydrate. *A. fumigatus* conidial suspension (2 μL, 3×10$^7$ CFU/mL) was later injected into the corneal stroma, and the left eye was used as a blank control group without treatment.[23] Processing is in accordance with the Association for Research in Vision and Ophthalmology (ARVO) Statement for the Use of Animals in Ophthalmic and Vision Research.

Mice with fungal keratitis were treated by subconjunctival injection of 5 μL of ZIF-90 (64 and 128 μg/mL) or DMSO (0.1%) once daily for 5 days. The progression of keratitis in mice was recorded with a camera under a slit lamp and scored clinically according to a 12-point scoring guideline.

**Hematoxylin eosin staining (HE staining)**

Eyeballs of mice infected with fungal keratitis were taken after 3 days of infection. The eyeballs were fixed for 48 h and then embedded in paraffin and sectioned (thickness of 10 μm. The sections were stained with hematoxylin and eosin stains. Then, the sections were photographed and recorded under a light microscope (Axio Vert; Zeiss, Jena, Germany, 400×).[24]

**Immunofluorescence（IF）**

The eyeballs of mice were taken and placed in an OCT embedding agent. The slices (10 μm) of the eyeballs were cut by freezing microtome upon the completion of quick freezing with liquid nitrogen. The slices were attached to adhesion pathology slides and roasted in an incubator at 37 °C for 8 h. Then, the specimens were fixed with methanol. Once fixation, it was soaked with PBS. The sample was dried and sealed with goat serum. Then, the specimen was incubated with NIMP-R14 antibody (1:100, Invitrogen) overnight (4 °C). The tablets were incubated for 1 h with the second antibody (FITC-conjugated) and finally sealed with anti-fluorescence attenuation tablets after 10 min of DAPI incubation. Images were photographed under fluorescence microscope (400x).

**Quantitative Real-Time PCR Experiments （RT-PCR)**

RAW 264.7 cells were inoculated in 12-well plates. When cells grew to 80%-90% of the well plate area, they were stimulated with inactivated *A. fumigatus* filaments for 1 h. The cells were treated with ZIF-90 (64 μg/mL) for 8 h. The cells were treated with ZIF-90 (64 μg/mL) for 8 h. Total ribonucleic acid (RNA) was collected to check for inflammatory factors IL-6, TNF-α, and IL-1β.

The corneas of mice at 3-day and 5-day postinfection (p.i.) were collected. Total RNA was collected, and inflammation-related factors IL-6, IL-1β, and TNF-α were examined by the same procedure as above. The inflammatory factors primers sequence of mice can be found in Table 1.

Table 1. Nucleotide sequences of mice primers for RT-PCR.

| Gene | GenBank No. | Primer Sequence (5' - 3') |
|---|---|---|
| GAPDH | NM_008084.2 | F: AAATGGTGAAGGTCGGTGTG |
|  |  | R: TGAAGGGGTCGTTGATGG |
| IL-1β (mouse) | NM_008361.3 | F: CGC AGC AGC ACA TCA ACA AGA GC |
|  |  | R: TGT CCT CAT CCT GGA AGG TCC ACG |
| *IL-6 (mouse)* | NM_001314054.1 | F: TGA TGG ATG CTA CCA AAC TGG A |
| | | R: TGT GAC TCC AGC TTA TCT CTT GG |
| *TNF-α (mouse)* | NM_013693.2 | F: ACCCTCACACTCAGATCATCTT |
| | | R: GGTTGTCTTTGAGATCCATGC |

**Enzyme-linked immunosorbent assay (ELISA)**

The supernatant of RAW 264.7 cells was taken and placed in PBS containing phenyl methyl sulfonyl fluoride (PMSF; Solarbio). Corneal tissue was crushed using a TissueLyser (QIAGEN, China). Next, the level of inflammatory protein in the supernatant was detected by ELISA kit (Biolegend, CA, USA). Protein concentration was determined by absorbance at 450 nm.

## 3. Results and discussion

### 1. Synthesis and characterization of ZIF-90

We successfully synthesized ZIF-90 with zinc nitrate and ICA (Figure 1A). The SEM characterization of ZIF-90 disclosed uniform and well-defined crystalline morphology (Figure 1B). The particles of ZIF-90 exhibited a regular shape and size distribution, indicating high-quality synthesis. The average size of ZIF-90 particles was 170 nm (Figure 1C). The XRD analysis further confirmed the crystalline structure of ZIF-90, displaying distinct diffraction peaks that align with the expected pattern of the ZIF-90 framework (Figure 1D). X-ray photoelectron spectroscopy (XPS) was employed to analyze the surface composition and chemical states of the elements present in ZIF-90. The survey spectrum revealed distinct peaks corresponding to zinc (Zn 2p, 1022.7 eV), carbon (C 1s, 248.8 eV), nitrogen (N 1s, 398.4 eV), and oxygen (O 1s, 531.1 eV), indicating the presence of ZIF-90 (Figure 1E). The high-resolution Zn 2p spectrum revealed characteristic peaks at approximately 1022 eV (Zn $2p_{1/2}$) and 1045 eV (Zn $2p_{3/2}$), confirming the coordination of zinc in the ZIF-90 framework (Figure 1F). The prominent peak around 1700 cm$^{-1}$ in the FT-IR spectrum indicates the presence of the aldehyde group (–CHO), which is a key feature distinguishing ZIF-90 from other ZIFs (Figure 1G). These results validated the successful synthesis of ZIF-90.[25]

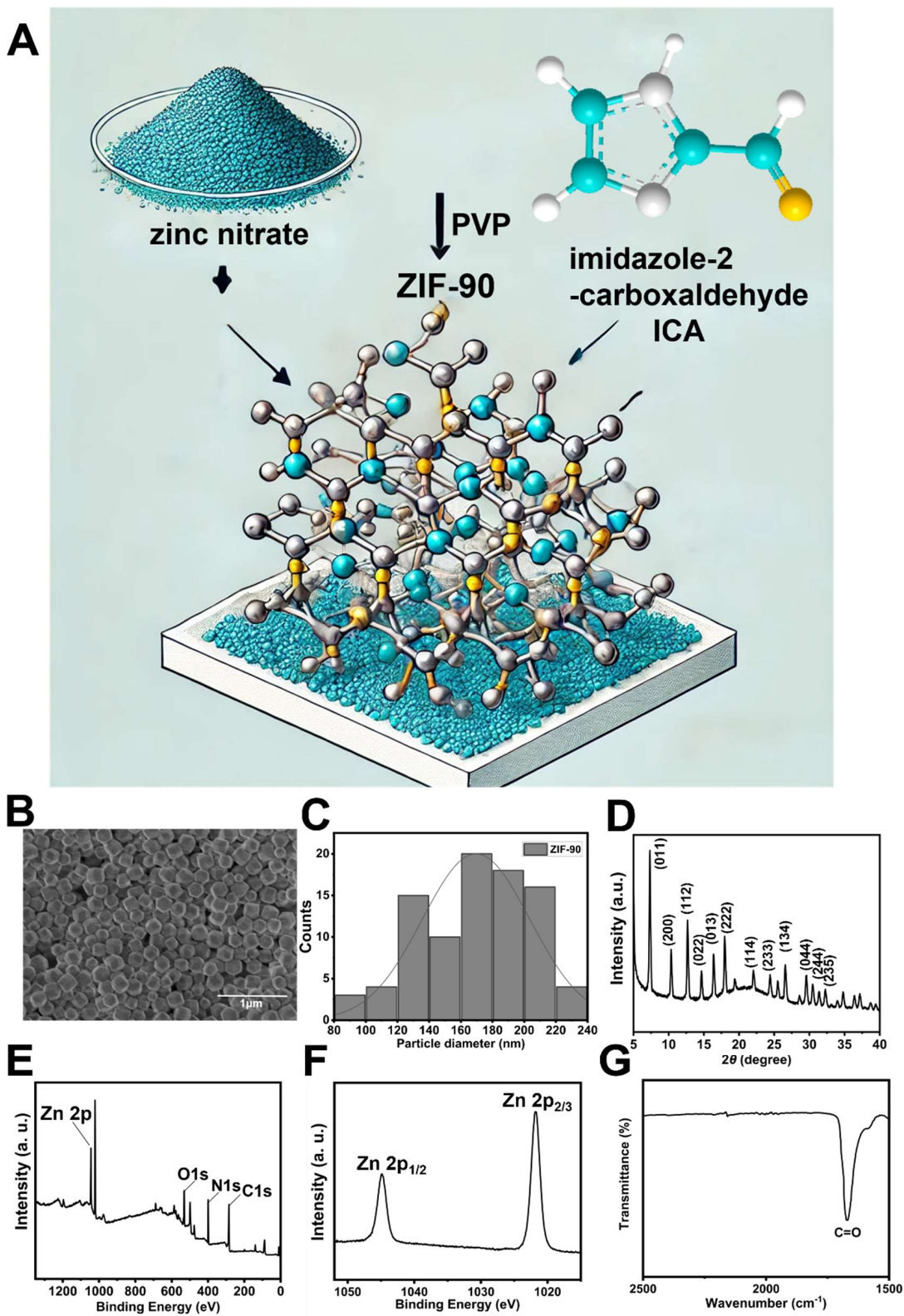

Fig. 1. Synthesis and characterization of ZIF-90 (A) Diagram of the synthesis process of ZIF-90 (B) SEM detection of ZIF-90 (Scale bar: 1μm) and (C) Particle size distribution graph of ZIF-90 (D) XRD pattern of ZIF-90 (E) XPS graphs (F)

**XPS graphs for Zn 2p (G) FT-IR maps of ZIF-90 can demonstrate the presence of aldehyde groups**

**2. Cytotoxicity in vitro and anti-*A. fumigatus* properties of ZIF-90**

In order to verify whether ZIF-90 can be used on HCECs and macrophages (RAW 264.7 cells), this test was conducted by cytotoxicity assay (CCK-8). The result of CCK-8 illustrated that ZIF-90 was non-toxic to both HCEC cells and RAW 264.7 cells at a concentration of 128 μg/mL or less (Figure 2A-B). The safe concentration of ZIF-90 for HCECs and RAW 264.7 cells in vitro is no more than 128 μg/mL

The spore adhesion experiment exhibited that ZIF-90 significantly reduced the adhesion of *A. fumigatus* spores to corneal epithelial cells (Figure 2C-D). The number of spores attached to the HCECs indicated that the number of spores in the ZIF-90 group was significantly lower than that in the control group (Figure 2E, P<0.0001). The conidia attached to the corneal epithelium then germinated into mycelia. This process releases toxins and enzymes, triggering an uncontrollable immune response and inflammation. These reactions may eventually lead to irreversible corneal damage.[26,27] ZIF-90 inhibited the adhesion of spores to corneal epithelial cells, which helps to reduce the occurrence of host immune response and inflammatory response, thereby reducing the irreversible damage to the cornea.

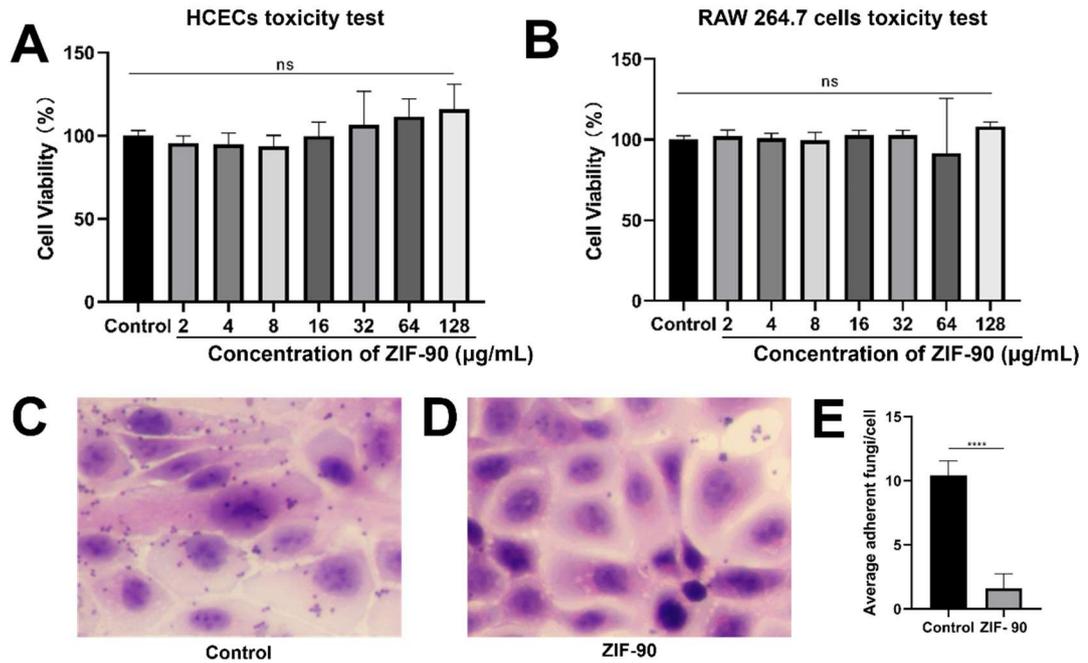

**Fig. 2. Cells cytotoxicity and anti-adhesion of *A. fumigatus* spores by ZIF-90. Effects of different ZIF-90 concentrations on (A) HCECs and (B) RAW 264.7 cells upon finishing incubation for 24 h, ZIF-90 (2-128 μg/mL). The adhesion of *A. fumigatus* to HCECs separately adding (C) DMSO and (D) ZIF-90. And (E) the statistical quantitative analysis of spore adhesion (ns, no significance, **** p<0.0001).**

To further explore the effect of ZIF-90 on *A. fumigatus*, we conducted a series of experiments, including spore adhesion test, MIC, biofilm inhibition test, PI, and calcium fluoride white staining. The MIC results displayed that ZIF-90 inhibited the formation of *A. fumigatus* at 32 μg/mL (Figure 3A, P<0.0001) and inhibited the formation of more than 90% *A. fumigatus* hyphae at the concentration of 64 μg/mL.

The biofilm inhibition experiment showed that ZIF-90 can suppress the biofilm formation of *A. fumigatus* at 16 μg/mL. More than 90% of biofilm formation was inhibited by ZIF-90 at 64 μg/mL (Figure 3B, P<0.0001). Biofilm formation is one of the main mechanisms of fungal resistance formation. The production of fungal cell biofilm played a crucial role in fungal resistance to clinical drugs. Inhibition of biofilm is conducive to preventing the formation of resistance to *A. fumigatus*.[28-30]

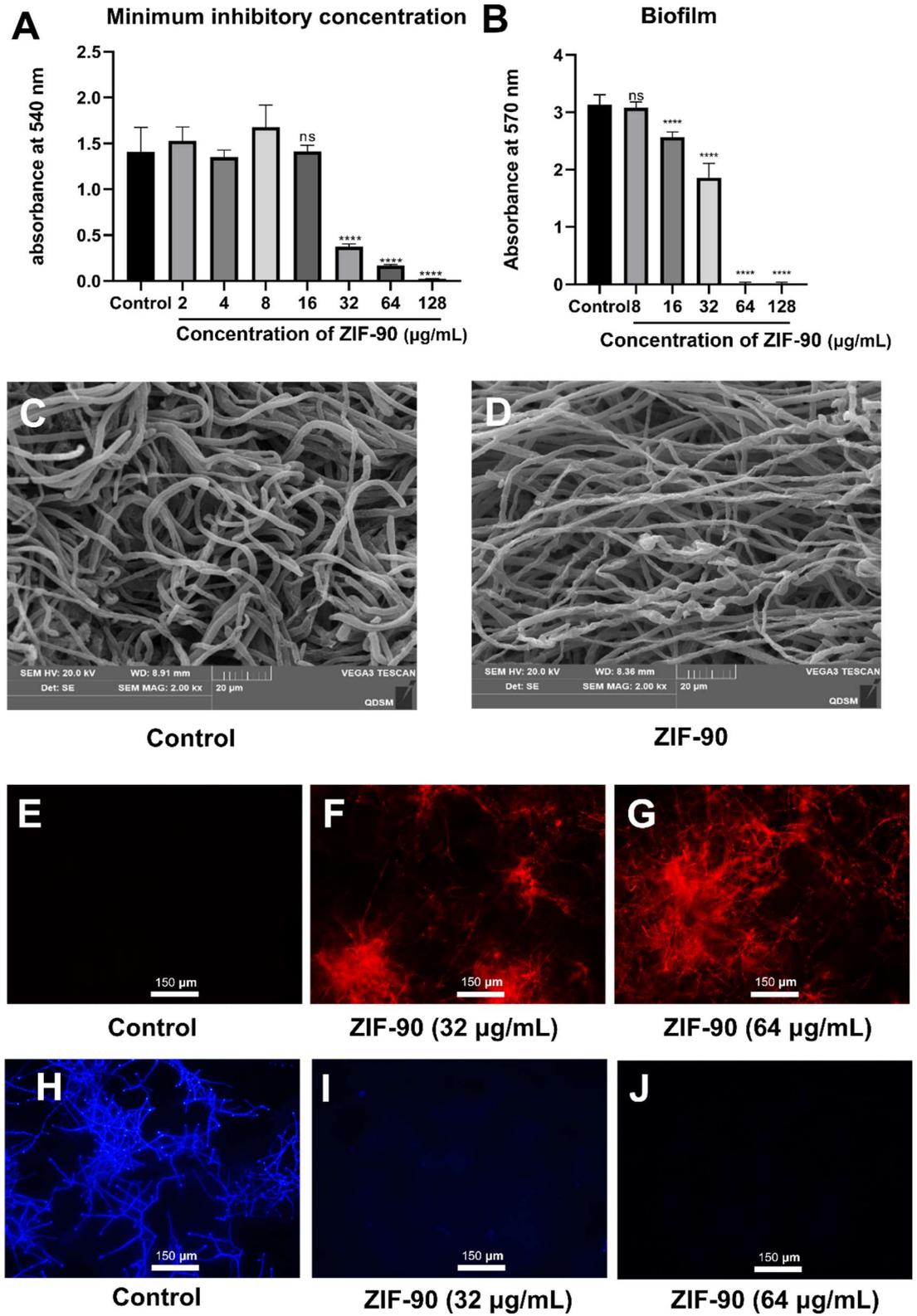

**Fig. 3. Anti-*A. fumigatus* action of ZIF-90** (A) Inhibiting the growth of *A. fumigatus* with different concentrations of ZIF-90, ZIF-90 (2-128 μg/mL). (B)

**Influence of ZIF-90 on inhibiting the growth of *A. fumigatus* biofilm, ZIF-90 (8-128 μg/mL). (C-D) Effect of ZIF-90 on mycelial morphology and structure of *A. fumigatus*, ZIF-90 (64 μg/mL). Impact of various ZIF-90 concentrations on the fungal cell membrane. (E-G) integrity and cell wall. (H-J) structure, ZIF-90 (32 and 64 μg/mL). (ns, no significance, **** p<0.0001)**

The SEM (Figure 3C-D) presented that the addition of ZIF-90 (Figure 3D) destroyed the fluency and integrity of *A. fumigatus* hyphae in comparison to the hyphae that only added DMSO (Figure 3C). PI staining indicated that compared with the control group (Figure 3E), the hyphae membrane of *A. fumigatus* in the ZIF-90 group was significantly damaged (Figure 3F-G). Calcium fluoride white stain was used to assess the changes in the cell wall (Figure 3H-J). Compared with the control group (Figure 3H), no fungal hyphal cell wall was observed after adding different concentrations of ZIF-90 (Figure 3I-J). Fungal cell walls and membranes are crucial in maintaining cell homeostasis and protecting cell stability. The above experiments demonstrated that ZIF-90 affected the homeostasis of fungal cells.

Antifungal and biofilm-inhibitory mechanisms of zinc ions ($Zn^{2+}$) and imidazole groups in ZIF-90 exhibit several noteworthy effects. Firstly, $Zn^{2+}$ ions exert antifungal activity by disrupting zinc-dependent enzymatic functions within fungal cells, leading to metabolic inhibition and cell death. Additionally, $Zn^{2+}$ ions compromise fungal cell membrane integrity, increasing permeability and causing leakage of vital cellular components. Moreover, $Zn^{2+}$ induces oxidative stress by generating reactive oxygen species (ROS), which damage critical cellular structures, including lipids, proteins, and DNA.[31]

ZIF-90 serves a key function in inhibiting fungal biofilm formation. They interfere with the synthesis of the extracellular matrix, a key component of biofilms, and disrupt quorum sensing, which is essential for biofilm development. Furthermore, imidazole groups may integrate into fungal cell membranes, altering their fluidity and reducing the ability of fungal cells to adhere to surfaces, thus preventing biofilm formation.[18,32,33]

The synergistic effect of zinc ions and imidazole groups within ZIF-90 significantly enhances antifungal efficacy by targeting multiple cellular mechanisms simultaneously. This combined action disrupts key processes within fungal cells, leading to improved antifungal performance. These findings illustrate the potential of ZIF-90 as a promising candidate for antifungal therapies, particularly in preventing biofilm-associated infections.

**3. ZIF-90 promotes apoptosis of macrophages and reduces inflammation**

Annexin V-FITC/PI apoptosis staining and Hoechst apoptosis staining were two classical methods for examining apoptosis. We verified the effect of ZIF-90 on the apoptosis of RAW 264.7 cells.

Annexin V is a calcium-dependent phospholipid-binding protein with a high affinity for phosphatidylserine (PS). The combination of PI and Annexin V can distinguish cells at different stages of apoptosis. Annexin V-FITC and PI double-positive cells are apoptotic cells. The results manifested that compared with the control group, treatment with hyphae stimulation alone, hyphae stimulation plus DMSO, and the addition of ZIF-90 (32 μg/mL) on the completion of hyphae stimulation did not promote the apoptosis of macrophages. The treatment with ZIF-90 (64 μg/mL; Figure 4A-E) significantly increased (Figure 4F, P<0.01) the number of apoptotic macrophages, proving that ZIF-90 can promote macrophage apoptosis.

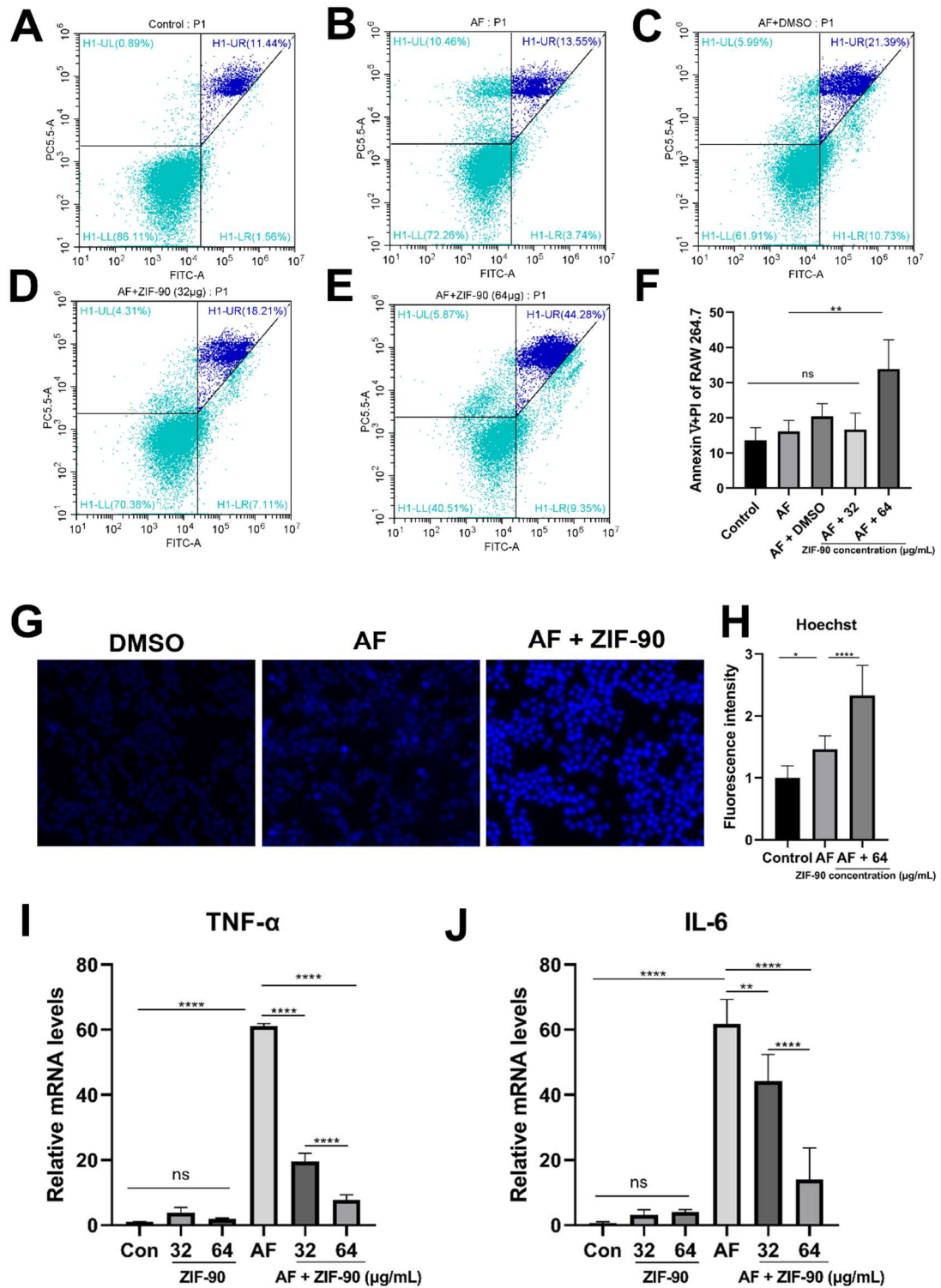

Fig. 4. ZIF-90 affects RAW 264.7 cells apoptosis. (A-E) Flow cytometry depicted that ZIF-90 (64 μg/mL) could promote the apoptosis of RAW 264.7 cells post *A. fumigatus* infection labeled with Annexin V and PI. (F) Statistical analysis of RAW 264.7 cells apoptosis rate dyed by Annexin V and PI. Hoechst staining and **statistical quantitative analysis of RAW 264.7 cells (G) showed that ZIF-90 (64 μg/mL) could promote apoptosis (H). The changes of pro-inflammatory factors TNF-α (I) and IL-6 (J). (ns, no significance; * p<0.05, ** p<0.01, **** p<0.0001)**

When cells undergo apoptosis, chromatin shrinks. Hoechst dye can penetrate the cell membrane, and the fluorescence of apoptotic cells is markedly enhanced than that of normal cells following staining. Our results highlighted that compared with the fungal stimulation alone group (AF, Figure 4G), the fluorescence of the group supplemented with ZIF-90 in succession to fungal stimulation was significantly enhanced (ZIF-90, Figure 4G). Besides, we conducted quantitative and statistical analysis of different groups and found that the fluorescence intensity of apoptotic macrophages in the ZIF-90 treatment group was indeed higher than that in the AF group (Figure 4H, p<0.0001). This result further confirms that ZIF-90 can promote macrophage apoptosis. In addition, through the changes of inflammatory factors, we can see that when the concentration of ZIF-90 increases from 32 μg/mL to 64 μg/mL, the contents of pro-inflammatory factors TNF-α (I) and IL-6 (J) are significantly decreased (P<0.0001) and the anti-inflammatory effect is significantly increased. These results suggest that ZIF-90 can further control inflammation by promoting apoptosis of macrophages.

Macrophages have a significant impact on the immune response to fungal keratitis by producing pattern-recognition receptors to clear the invading fungus. Excessive immune and inflammatory responses can damage the corneal epithelium and decrease vision.[5,7,34] Hence, the regulation of macrophage survival and its production of inflammatory factors are pivotal in the protection of corneal cells. Apoptosis is a regulated and controlled process that contributes importantly in maintaining homeostasis and tissue development by eliminating damaged, unnecessary, or potentially harmful cells without triggering an inflammatory response. Besides, it has been demonstrated that apoptotic macrophages can release specific anti-inflammatory substances while maintaining the integrity of the plasma membrane.[35] Thus, we investigated the effect of ZIF-90 on macrophage apoptosis. Accordingly, it can be concluded that ZIF-90 can promote apoptosis of macrophages, and the effect of ZIF-90 on reducing inflammatory factors is through promoting apoptosis of macrophages.

**4. Verification of corneal toxicity of ZIF-90 in mice**

The important prerequisite for the application of the drug in the eyes of mice is that the drug is non-toxic to the corneal epithelium of mice. The results presented that no corneal defect was observed under cobalt blue light at 0, 1, 3, and 5 days after the eye subconjunctival injection (Figure 5A). Additionally, the score of corneal fluorescein sodium staining was found to be 0 score both in the control group and ZIF-90 treatment group at different concentrations (64 and 128 μg/mL) at days 0, 1, 3, and 5 (Figure 5B). To summarize, ZIF-90 of 64 and 128 μg/mL had no potential toxicity to the corneal epithelium of mice. The results manifest that ZIF-90 was non-toxic to the ocular surface of mice and could be used to treat fungal keratitis in mice.

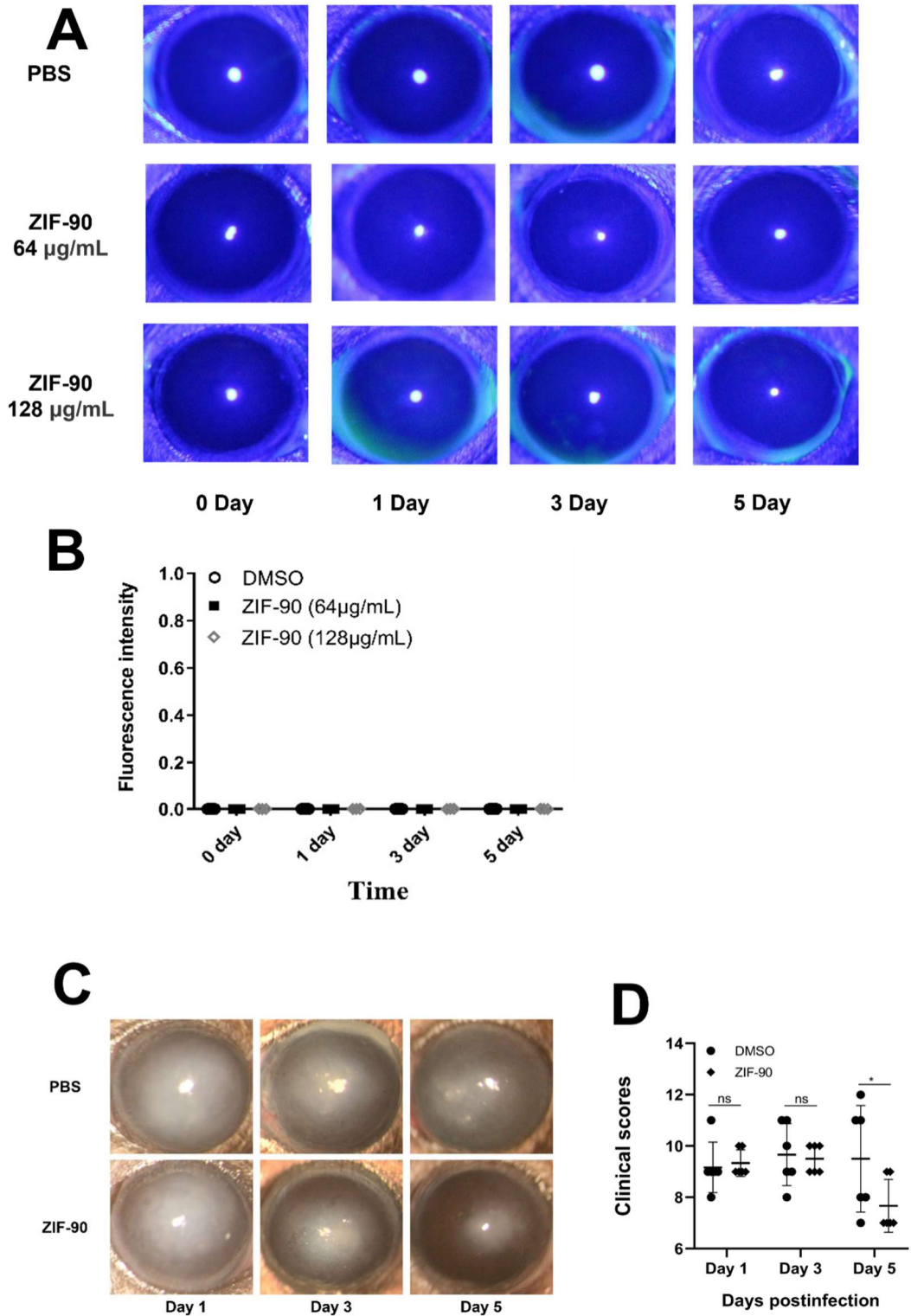

Fig. 5. Corneal toxicity of ZIF-90 and therapeutic effect of ZIF-90 on fungal keratitis in mice. Fluorescein sodium staining of mice cornea using ZIF-90 (A) and fluorescence intensity (B) succeeding staining. Therapeutic effect (C) and clinical score (D) of ZIF-90 in mice with fungal keratitis. (ns, no significance; *

p<0.05)

**5. ZIF-90 for the treatment of mouse *A. fumigatus* keratitis**

Ocular anterior segment photographs of mice with fungal keratitis were captured using a slit lamp. In the PBS control group, corneal opacity, ulceration, and epithelial defects were evident at 3 days of *A. fumigatus* infection, with an escalation in the severity of the infection observed in the corneas later on 5 days (Figure 5C). The corneas treated with ZIF-90 exhibited slight structural damage and a reduction in opacity. Additionally, clinical evaluations demonstrated significantly lower scores (Figure 5D) in the ZIF-90 group, indicating improved therapeutic outcomes. Therefore, it is established that ZIF-90 has the effect of treating fungal keratitis.

Imidazole derivatives, which form the core structure of various antifungal agents (e.g., ketoconazole and miconazole), exert their effects by inhibiting the biosynthesis of ergosterol. Disruption of ergosterol compromises membrane integrity could lead to increased permeability and eventual fungal cell lysis. Zinc ions ($Zn^{2+}$) also contribute to antifungal activity by interfering with fungal metabolic processes, inhibiting key enzymes involved in fungal growth, and destabilizing fungal biofilm structures. Furthermore, zinc ions impede fungal spore germination and hyphal development, thereby preventing fungal colonization.[36-38] This makes them a promising strategy of ZIF-90 for treating persistent fungal infections, including fungal keratitis.

**6. Corneal cell inflammatory changes and Anti-inflammatory effect of ZIF-90**

The control of inflammatory response is a crucial indicator for the effective treatment of fungal keratitis. The downregulation of inflammatory cell infiltration can imply the inflammatory response is well controlled. HE staining showed that ZIF-90 decreased the content of inflammatory cells in corneal tissues infected with fungal keratitis (Figure 6A). Neutrophils are the most abundant inflammatory cells in the early stage of fungal keratitis. Immunofluorescence staining denoted the content of neutrophils in the cornea of mice infected with *A. fumigatus* after ZIF-90 treatment also decreased (Figure 6B). As well we verified that the pro-inflammatory factors IL-1β, IL-6, and TNF-α in the cornea of mice treated with ZIF-90 were significantly reduced on the 3 and 5 days p.i. compared with those treated with PBS (Figure 6C-E).

Excessive infiltration of inflammatory cells during the development of fungal keratitis contributes to significant corneal tissue damage. Neutrophils and macrophages release pro-inflammatory cytokines and reactive oxygen species (ROS), which not only target the fungal infection but also lead to corneal stromal degradation, ulceration, and opacity.[39] This results in impaired wound healing and, in severe cases, corneal perforation, causing vision impairment. Effective treatment requires careful management of the immune response to avoid exacerbating inflammation and tissue damage. Treatment with ZIF-90 effectively controlled the immune response and avoided increased inflammation and tissue damage.

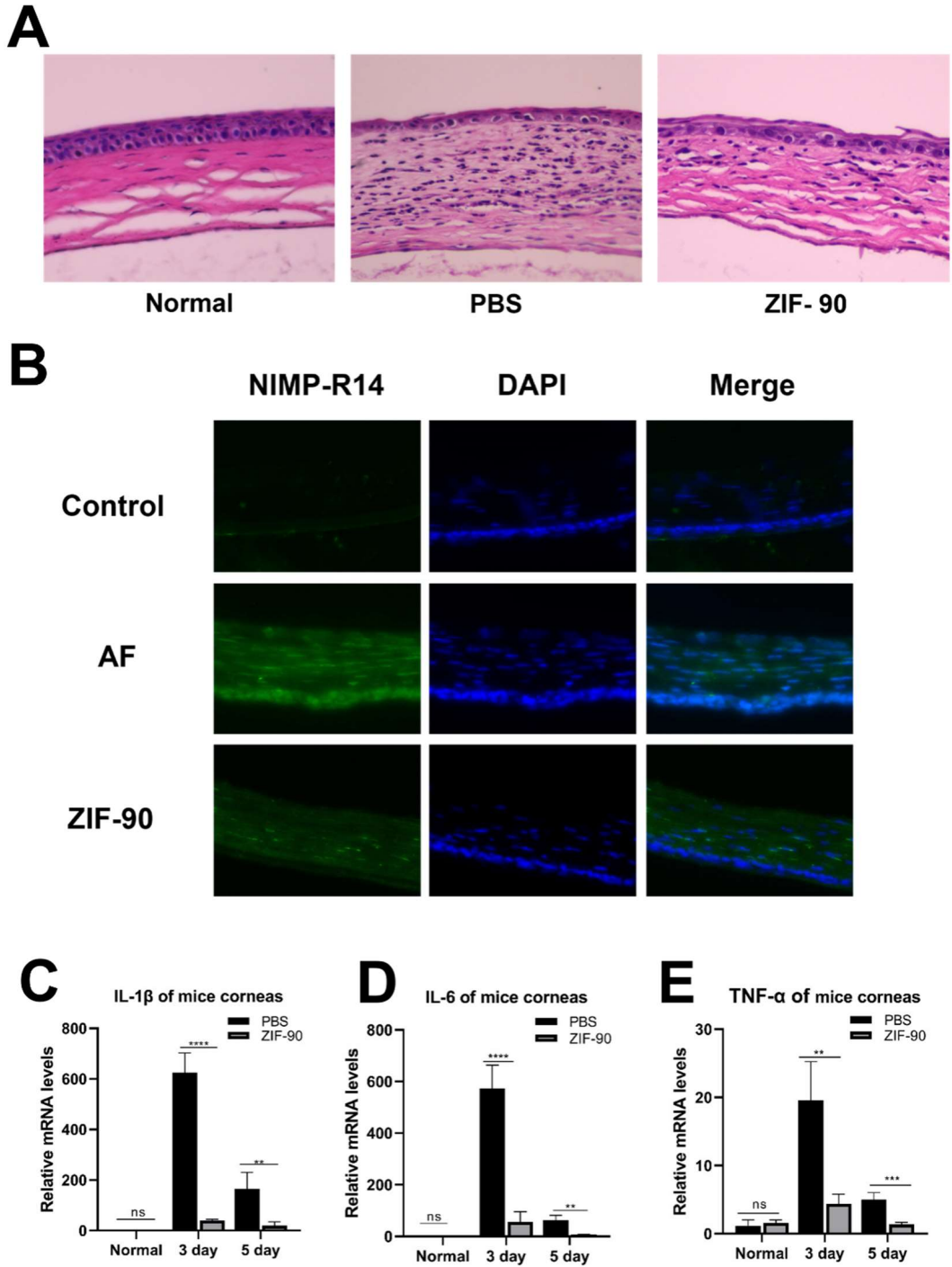

**Fig. 6. Changes of corneal inflammatory cells and inflammatory factors in mice (A) HE staining reflected the penetration of inflammatory cells in different groups of mice three days in line with infection. (B) Fluorescent staining of neutrophil infiltration in mice infected with fungal keratitis. Infiltration of corneal inflammatory factors (C) IL-1β, (D) IL-6, and (F) TNF-α in mice. (ns, no**

**significance; ** p<0.01, *** p<0.001, **** p<0.0001)**

## 7. Cytotoxicity, antifungal and anti-inflammatory effects of ZIF-8, ZIF-67 and MOF-74

In order to endorse the effective antifungal components in ZIF-90 and to judge the biocompatibility of ZIF-90, we identified ZIF-8, ZIF-67, and MOF-74 (Zn) with different metal ion compositions and different organic ligands from ZIF-90. ZIF-8 differs from ZIF-90 by an aldehyde group. ZIF-67 differs from ZIF-90 by an aldehyde group (-CHO) and zinc ions. MOF-74 (Zn) shares zinc ions with ZIF-90 but has different organic ligands. These three MOF materials are also widely used as carriers in the biomedical field.[40-44] While verifying the antifungal component of ZIF-90, it is also possible to certify the antifungal potential of other MOF materials to expand the application field of MOF materials.

To study the cytotoxicity of these materials, the cellular activity of HCECs and RAW 264.7 cells was detected by CCK-8 kit. The results indicated that ZIF-90 was non-toxic to both HCECs and RAW 264.7 cells at concentrations less than 128 μg/mL (Figure 2A-B). The other three materials were toxic to HCECs and RAW 264.7 cells at 64 μg/mL (Figure 7A-F), and MOF-74 (Zn) was toxic to RAW 264.7 cells at 32 μg/mL (Figure 7F, p<0.0001). Compared to ZIF-8, ZIF-67, and MOF-74 (Zn), ZIF-90 exhibited better biocompatibility, reflecting lower cytotoxicity. Since the difference between ZIF-90 and ZIF-8 was only in the structure of the aldehyde group and a hydroxyl group, we manifest that the presence of the aldehyde group is conducive to reducing the toxicity of ZIF materials.

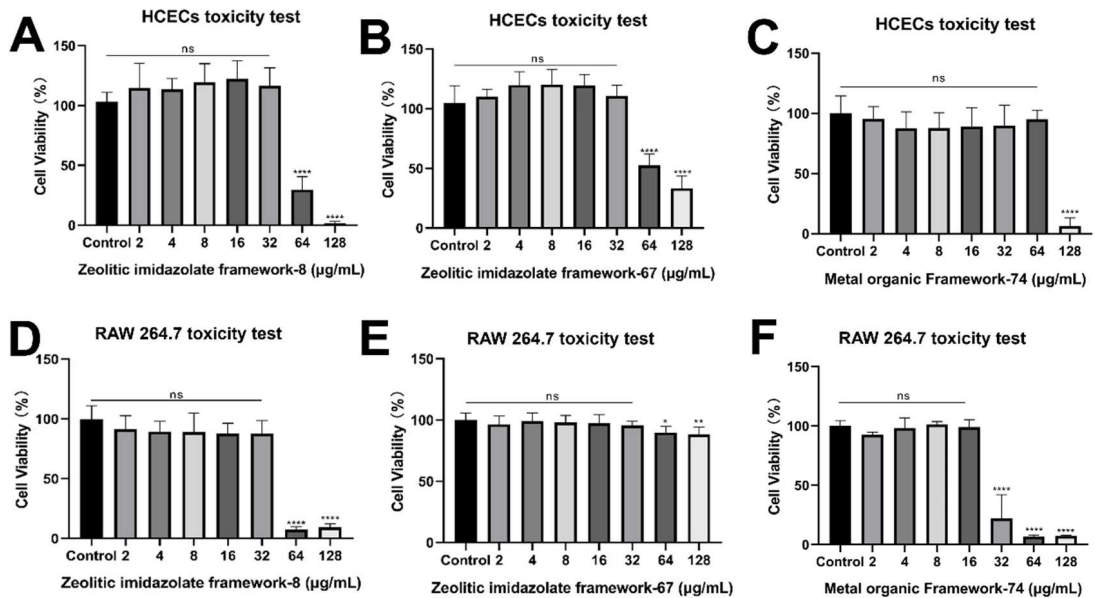

Fig. 7. Poisonousness of different MOF materials to different cells. Toxicity of materials at different levels of (A) ZIF-8, (B) ZIF-90, and (C) MOF-74 to HCECs. And Cytotoxicity of (A) ZIF-8, (B) ZIF-90, and (C) MOF-74 towards RAW 264.7 cells. (ns, no significance; * p<0.05, ** p<0.01, **** p<0.0001)

MIC experiment results illustrated that MOF-74 (Zn) has the ability to block the growth of *A. fumigatus* at 16 μg/mL (Figure 8A, p<0.0001). At 32 μg/mL, ZIF-8, ZIF-67, MOF-74 (Figure 8A-C, p<0.0001) and ZIF-90 (Figure 3A) all inhibited fungal growth. When the concentration reached 64 μg/mL, ZIF-8, MOF-74 (Figure 8A-B), and ZIF-90 (Figure 3A) could impede more than 80% of the fungal growth. The antifungal rate of these three materials exceeded 90% at 128 μg/mL. However, the inhibition rate of ZIF-67 could not reach 80% even at a concentration of 128 μg/mL (Figure 8C, p<0.0001). Although these four MOF materials have good anti-fungal effects, the effect of ZIF-67 is not as good as the other three materials. In these four MOF materials, ZIF-67 is the only MOF material whose metal ions are cobalt ions rather than zinc ions. It can be seen that zinc ions in MOF materials have certain anti-fungal effects. And MOF materials containing imidazole groups and MOF-74 (Zn) containing heterocyclic acids have anti-fungal effects. In addition, under the premise of non-toxic to cells, the anti-fungal effect of ZIF-90 is stronger than the other three materials.

Furthermore, we verified the effect of four materials on reducing pro-inflammatory factors in RAW 264.7 cells. We found that all four materials had the effect of reducing pro-inflammatory factors TNF-α and IL-6 at the protein level. Additionally, ZIF-8, MOF-74, and ZIF-90 had better effects on minimizing the pro-inflammatory protein TNF-α than ZIF-67 (Figure 8 D-E, p<0.0001). The experimental results confirmed that ZIF-8, ZIF-67, MOF-74, and ZIF-90 can lower the pro-inflammatory factors TNF-α and IL-6, so it is beneficial to decrease the inflammatory response.

According to the experimental results, zinc and imidazole ligands in ZIF-90 can jointly exert anti-inflammatory and antifungal effects. This expanded the previous studies on the antibacterial effect of MOF materials containing heterocyclic acid ligands[45] and verified that MOF materials contain imidazole ligands. These outcomes develop the antifungal function of MOF materials.

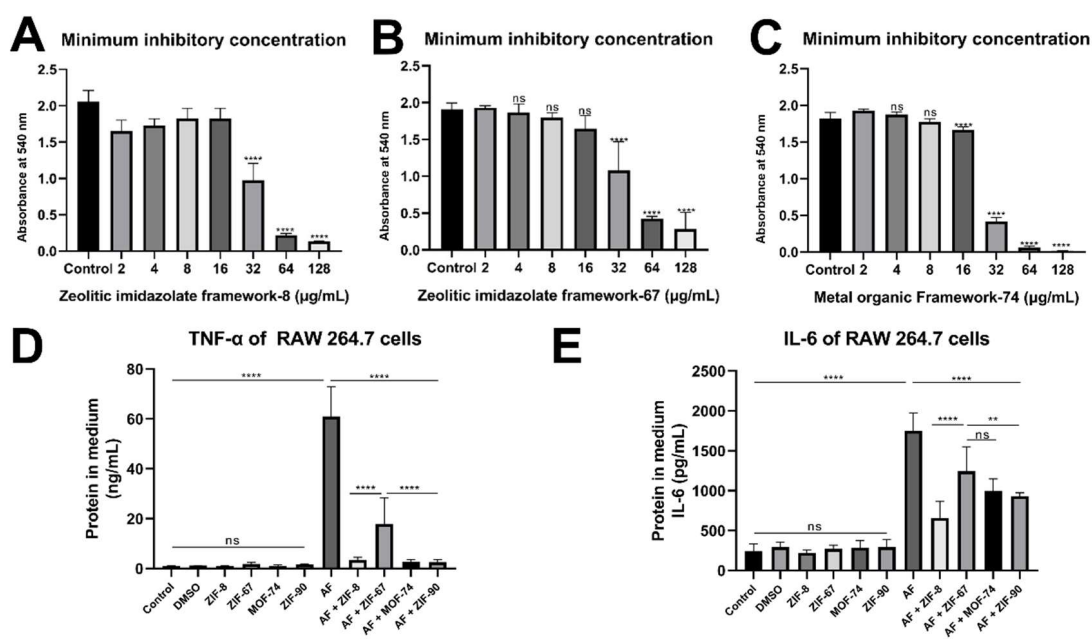

**Fig. 8. The outcome of antifungal and anti-inflammatory of different MOF materials. Anti- *A. fumigatus* activity of (A) ZIF-8, (B) ZIF-67, and (C) MOF-74 at varying dosages*.* Effect of ZIF-8, ZIF-67, and MOF-74 on reducing proinflammatory factor content (D) TNF-α and (E) IL-6 in RAW 264.7 cells. (ns, no significance; ** p<0.01, **** p<0.0001)**

## 4. Conclusion

The effect of ZIF-90 on fungal keratitis was studied in this paper. ZIF-90 can hamper the growth of *A. fumigatus*. ZIF-90 can also prevent excessive immune response and inflammatory response by promoting macrophage apoptosis. In addition, ZIF-90 can treat fungal keratitis by reducing corneal scores and inflammatory cell infiltration. Both zinc ions and imidazole ligands of ZIF-90 have anti-inflammatory and antibacterial effects. Moreover, the presence of an aldehyde group on the imidazole ligand enhanced the biocompatibility of ZIF-90. These results suggest that ZIF-90 has good anti-inflammatory properties and immunogenicity, which provides a basis for the treatment of fungal keratitis.

## Acknowledgments

This work was financially supported by China Postdoctoral Science Foundation (Nos. 2020M672000) and the Taishan Scholars Program (Nos. ts201511108, tsqn202103188 and tsqn201812151).